\documentclass{article}
\usepackage{spconf,amsmath,graphicx,hyperref}

\title{FOA Tokenizer: Low-Bitrate Neural Codec for First Order \\ Ambisonics with Spatial Consistency Loss}
\twoauthors
  {Parthasaarathy Sudarsanam\sthanks{Work completed during internship at Microsoft.}}
	{Tampere University \\ 
    Tampere, Finland}
  {Sebastian Braun, Hannes Gamper}
	{Microsoft Research\\
    Redmond, USA}

\begin{document}
\ninept
\maketitle
\begin{abstract}
Neural audio codecs have been widely studied for mono and stereo signals, but spatial audio remains largely unexplored. We present the first discrete neural spatial audio codec for first-order ambisonics (FOA). Building on the WavTokenizer architecture, we extend it to support four-channel FOA signals and introduce a novel spatial consistency loss to preserve directional cues in the reconstructed signals under a highly compressed representation. Our codec compresses 4-channel FOA audio at 24 kHz into 75 discrete tokens per second, corresponding to a bit rate of 0.9 kbps. Evaluations on simulated reverberant mixtures, non-reverberant clean speech, and FOA mixtures with real room impulse responses show accurate reconstruction, with mean angular errors of 13.76°, 3.96°, and 25.83°, respectively, across the three conditions. In addition, discrete latent representations derived from our codec provide useful features for downstream spatial audio tasks, as demonstrated on sound event localization and detection with STARSS23 real recordings.
\end{abstract}
\section{Introduction}
\label{sec:intro}


Spatial audio captures the way humans perceive sound in three-dimensional space, offering a natural and immersive auditory experience. It plays a central role in emerging technologies such as virtual and augmented reality, gaming, and next-generation media streaming, where the sense of spatial presence is essential. As these applications grow in scale and complexity, there is an increasing demand for methods that can represent and process spatial audio efficiently, without compromising perceptual quality.


Spatial audio is often represented using First-Order Ambisonics (FOA), which encodes directional information in a compact, spherical format suitable for processing and reproduction. Learning a compressed discrete representation of FOA-based spatial audio is a key step toward enabling efficient transmission, spatial audio understanding, spatial audio language models, and generative modeling. Compressed representations are important for 
efficient and bandwidth-constrained data transmission and storage, as well as efficient generative modeling such as LLMs and latent diffusion models.
At the same time, discrete latent spaces align naturally with audio-language models that operate on tokenized inputs. Such representations provide a foundation for spatial audio synthesis and support downstream tasks such as sound source localization.

\begin{figure}
    \centering
    \includegraphics[width=0.8\linewidth]{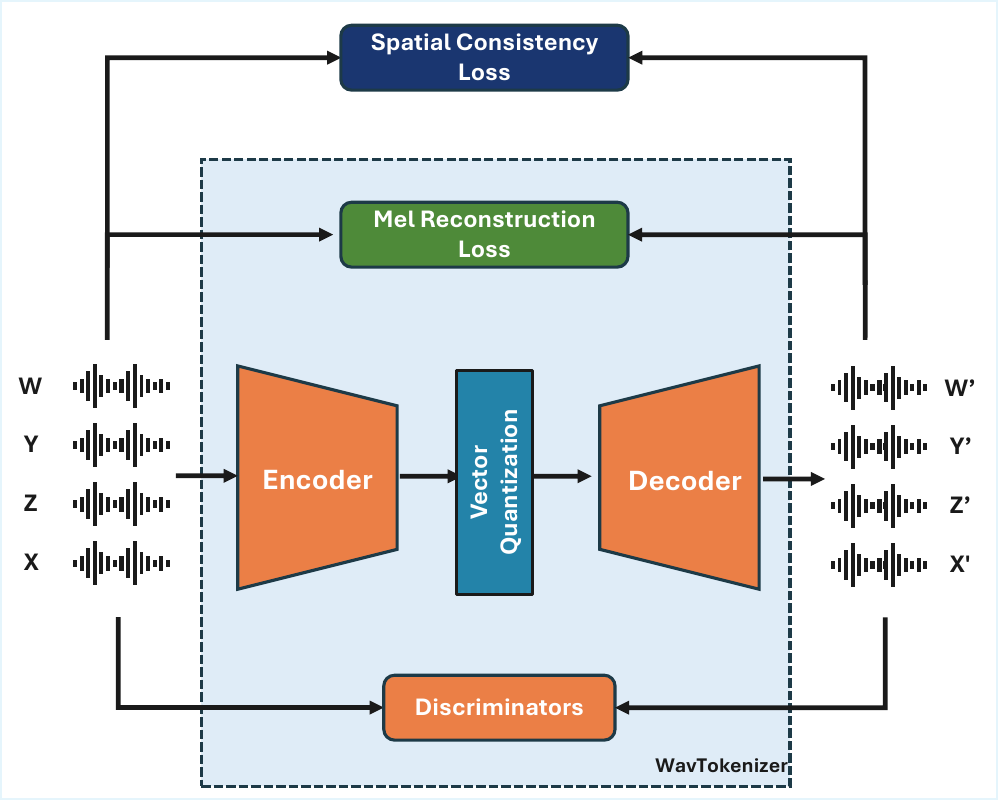}
    \caption{Proposed FOA spatial audio codec with spatial consistency loss.}
    \label{fig: architecture}
\end{figure}

Neural audio codecs, such as SoundStream \cite{zeghidour2021soundstream}, Encodec~\cite{defossez2022highfi}, and DAC \cite{kumar2023high}, achieve high-fidelity reconstruction of single-channel audio at low bitrates. These models adopt a U-Net based encoder-decoder architecture and use residual vector quantization (RVQ) at the bottleneck to produce discrete representations. While RVQ enhances reconstruction quality, it increases the number of discrete tokens and poses challenges for generative modeling. Recently, WavTokenizer \cite{ji2025wavtokenizer} achieved high-quality reconstruction using a single broader VQ layer. The model compresses 24 kHz audio into 75 discrete tokens per second while achieving performance comparable to RVQ-based methods at lower bitrates, providing an efficient approach for audio compression. 

Representation learning for FOA has been an active area of research. Methods such as ELSA \cite{spatially-aware} and MC-SimCLR \cite{jiang2024exploring} use contrastive learning techniques to learn spatial audio representations. More recently, ImmerseDiffusion \cite{heydari2025immersediffusion} and SonicMotion \cite{templin2025sonicmotion} adopt convolutional U-Net architectures similar to DAC, replacing RVQ layers with continuous VAE-based representations, achieving a 128$\times$ compression.

In this work, we extend the single-channel WavTokenizer to develop the first discrete neural spatial audio codec. We evaluate the performance of our FOA codec across a variety of datasets to evaluate both the acoustic and spatial reconstruction quality. Our contributions are two-fold:
\begin{itemize}
    \item We present the first vector-quantized representation for 4-channel FOA spatial audio, using 75 tokens per second at 24 kHz, achieving 320x compression and an effective bandwidth of 0.9 kbps.
    \item We propose a novel spatial consistency loss that enables the preservation of directional characteristics of sound events under this highly compressed representation.
\end{itemize}



\section{METHODOLOGY}

\subsection{Spatial audio codec}

Our proposed spatial audio codec is shown in Figure \ref{fig: architecture}. Our design builds upon the WavTokenizer framework \cite{ji2025wavtokenizer}, which compresses single-channel audio into discrete tokens through an encoder, a single VQ layer, and an asymmetric Vocos decoder \cite{siuzdak2024vocos}. WavTokenizer employs strided convolutional blocks in the encoder to downsample the audio 320x to generate latent representations, a codebook with 4096 codes to discretize the latents, a decoder based on ConvNeXt blocks and attention modules, and iSTFT to reconstruct the waveform.

To extend this framework to spatial audio, we modify the encoder’s first convolutional layer to accept four-channel first-order ambisonics (FOA) audio sampled at 24 kHz. The encoder then compresses the FOA signal into a latent representation using the same architecture as in WavTokenizer. The vector quantization layer maps these latents into a discrete codebook space with 4096 entries of dimension 512, initialized via K-means clustering and updated with an exponential moving average. To prevent codebook collapse, we also apply a dead code reactivation strategy \cite{dhariwal2020jukeboxgenerativemodelmusic}, reinitializing unused codes by sampling latent vectors from the current batch. The decoder reconstructs the waveform from the quantized latent representations. It employs 
the same decoder backbone as the WavTokenizer, with the final iSTFT head producing four channels of the reconstructed FOA audio.

To improve perceptual quality, our spatial audio codec is trained with a series of discriminators that take the four-channel FOA input or reconstructed signals. Specifically, we employ a multi-period discriminator, a multi-resolution STFT discriminator, and a DAC discriminator, which encourage the reconstructed waveform to match the input in both temporal and spectral structures. This setup establishes our FOA-VQGAN framework capable of high-fidelity spatial audio reconstruction.

\subsection{Spatial consistency loss}

We introduce a novel spatial consistency loss ($\mathcal{L}_{sc}$) inspired by the principles of Directional Audio Coding (DirAC)~\cite{e536cacb04f141eb9179c6628c23290c}, which models spatial perception in terms of energy, diffuseness, and intensity vector directions. The $\mathcal{L}_{sc}$ compares the intensity vectors extracted from the FOA representation of the input and reconstructed signals. The directional agreement is quantified using cosine similarity

\begin{equation}
s_{t,k} = \cos\theta_{t,k} \;=\;
\frac{\mathbf{I}^{(i)}_{t,k}\cdot\mathbf{I}^{(r)}_{t,k}}
{\lVert\mathbf{I}^{(i)}_{t,k}\rVert_2\,\lVert\mathbf{I}^{(r)}_{t,k}\rVert_2+\varepsilon},
\end{equation}

\noindent where $\mathbf{I}^{(i)}_{t,k}$ and $\mathbf{I}^{(r)}_{t,k}$ denote the intensity vectors at time $t$ and frequency bin $k$ for the input and reconstructed signals, respectively.

As spatial direction is reliable only in regions dominated by energetic and non-diffuse sources, we introduce a binary mask that discards low-energy and highly diffuse components given by
\begin{equation}
m_{t,k} = \mathbf{1}\{E^{(i)}_{t,k}>\tau_E\}\,\mathbf{1}\{D^{(i)}_{t,k}<\tau_D\},
\end{equation}

where $E^{(i)}_{t,k}$ and $D^{(i)}_{t,k}$ denote the energy and diffuseness of the input signal at time $t$ and frequency $k$, with thresholds $\tau_E$ and $\tau_D$. To emphasize regions with strong and clear directional cues, each region is weighted by its energy and by how little diffuseness it contains,

\begin{equation}
w_{t,k} = m_{t,k}\,E^{(i)}_{t,k}\,(1-D^{(i)}_{t,k}).
\end{equation}

The spatial consistency loss $\mathcal{L}_{sc}$ is defined as the weighted penalty on misaligned intensity vectors:
\begin{equation}
\mathcal{L}_{sc} = \frac{1}{TK}\sum_{t=1}^{T}\sum_{k=1}^{K} 
w_{t,k}\,\big(1-s_{t,k}\big).
\end{equation}

This term is added to the overall generator objective, ensuring that spatial alignment is optimized alongside the other losses. The overall loss used to train the generator is given by

\begin{equation}
\begin{aligned}
\mathcal{L}_{gen} &= \lambda_q \mathcal{L}_q + \lambda_{mel} \mathcal{L}_{mel} + \lambda_{adv} \mathcal{L}_{adv} \\
&\quad + \lambda_{feat} \mathcal{L}_{feat} + \lambda_{sc} \mathcal{L}_{sc}
\end{aligned}
\end{equation}

\noindent where $\mathcal{L}_q$ is the quantization loss, $\mathcal{L}_{mel}$ the mel-spectrogram reconstruction loss, $\mathcal{L}_{adv}$ the adversarial loss, $\mathcal{L}_{feat}$ the feature-matching loss, and $\mathcal{L}_{sc}$ the spatial consistency loss. The coefficients $\lambda_q,\lambda_{mel},\lambda_{adv},\lambda_{feat},\lambda_{sc}$ control their relative weights. 

The discriminators are trained with the standard hinge loss given by,

\begin{equation}
\begin{aligned}
\mathcal{L}_{dis}(X,\tilde{X}) &= \frac{1}{K}\sum_{k=1}^{K} 
\Big( \max(0,1-D_k(X)) \\
&\quad + \max(0,1+D_k(\tilde{X})) \Big),
\end{aligned}
\end{equation}

\noindent where $D_k(\cdot)$ is the $k$-th discriminator output, $X$ the input FOA signal, and $\tilde{X}$ the reconstructed FOA signal, providing a stable adversarial signal to guide the generator.

\begin{table*}[!ht]
\centering
\vspace{-6pt} 
\caption{Comparison of acoustic and spatial reconstruction metrics for different codecs and datasets.}
\vspace{0.1cm}
\resizebox{\textwidth}{!}{%
\begin{tabular}{c|c|ccccc|ccc}
\hline
\textbf{Codec} & \textbf{Dataset} & \multicolumn{5}{c|}{\textbf{Acoustic Reconstruction}} & \multicolumn{3}{c}{\textbf{Spatial Reconstruction}} \\
\cline{3-7} \cline{8-10}
 &  & CLAP \(\uparrow\) & STFT Dist. \(\downarrow\) & Mel Dist. \(\downarrow\) & DistillMOS \(\uparrow\) & WER \(\downarrow\) & Azimuth Err. \(\downarrow\) & Elevation Err. \(\downarrow\) & Angular Err. \(\downarrow\) \\
\hline
Opus (24kbps) &  & 0.78 & 3.40 & 2.18 & - & - & 23.18° & 10.62° & 22.47° \\
Opus (32kbps) & In-domain & 0.84 & 3.34 & 1.98 & - & - & 7.39° & 4.62° & 8.06° \\
FOA-VQGAN (0.9kbps) &  & 0.92 & 1.60 & 1.28 & - & - & 11.20° & 9.07° & 13.76° \\
\hline
\hline
Opus (24kbps) &  & 0.93 & 2.74 & 1.93 &  2.42 & 0.23 & 18.98° & 7.61° & 17.23° \\
Opus (32kbps) & SpatialVCTK & 0.95 & 2.42 & 1.51 & 2.98 & 0.16 & 0.81° & 0.60° & 1.02° \\
FOA-VQGAN (0.9kbps) &  & 0.96 & 1.62 & 1.39 & 3.07 & 0.67 & 3.50° & 2.81° & 3.96° \\
\hline
\hline
Opus (24kbps) &  & 0.79 & 3.98 & 2.43 & - & - & 39.32° & 15.56° & 40.17° \\
Opus (32kbps) & MEIR  & 0.84 & 4.00 & 2.21 & - & - & 11.11° & 7.43° & 13,28° \\
FOA-VQGAN (0.9kbps) & & 0.88 & 1.82 & 1.43 & - & - & 17.23° & 18.42° & 25.83° \\
\hline
\end{tabular}%
}
\label{tab:main_results}
\end{table*}



\section{Evaluation}

\subsection{Dataset}


For training, we constructed a large-scale synthetic dataset consisting of  2 million 10-second recordings. Spatial FOA room impulse responses (RIRs) were simulated using the image-source method implemented in pyroomacoustics \cite{scheibler2018pyroomacoustics}, by generating 10k unique rooms and microphone positions with 64 source candidates per room, spherically uniformly distributed around the microphone. As audio material we used speech from CommonVoice \cite{ardila2019common} spanning 8 languages (385~h) and general audio from Freesound \cite{10.1145/2502081.2502245} ($\sim$230k files) and BBC Sound Effects\footnote{https://sound-effects.bbcrewind.co.uk/} ($\sim$33k files). General audio clips were divided into single source sounds ($\sim$700~h) and multi-source or ambient sounds ($\sim$4000~h) based on their tags and descriptions. The single source material and speech were spatialized in distinct directions, while the multi-source/ambient was used to generate diffuse background sounds by convolving with all 64 RIRs. We mixed randomly 1-5 stationary directional sources and optional diffuse background sound with varying levels to generate diverse acoustic characteristics. 

For evaluation, we considered three complementary datasets. First, 10k recordings were generated using the same simulation strategy as training, but with unseen audio sources from SoundBible\footnote{https://soundbible.com/} and different 1000 rooms to assess generalization to new content. We refer to this evaluation dataset as in-domain dataset. Second, a SpatialVCTK dataset was created by spatializing clean VCTK speech recordings \cite{yamagishi2019vctk}, without background noise or reverberation, to provide a controlled benchmark. Third, real measured RIRs from the MEIR dataset \cite{9747603} were combined with anechoic sounds and real spatial background noise recordings from MEIR to test robustness under realistic conditions.

\subsection{Training details}

We trained our codec for a total of 1M steps, comprising 500k steps each for the generator and the discriminators on a cluster of A100 GPUs with a batch size of $128$. We used the AdamW optimizer with a learning rate of $2 \times 10^{-4}$ and a cosine scheduler. The weight for the mel reconstruction loss ($\lambda_{\text{mel}}$) and the commitment loss ($\lambda_{\text{q}}$) were set to 45 and 1000, respectively, following the configuration used in WavTokenizer. For spatial consistency loss, we used $\lambda_{\text{sc}} = 1$, with an energy threshold of $10^{-6}$ and a diffuseness threshold of 0.95.

\section{Results}

In Table \ref{tab:main_results}, we compare the performance of our spatial audio codec (FOA-VQGAN) against multichannel Opus \cite{rfc6716} at various bitrates.  For acoustic reconstruction, we report the CLAP similarity score between input and reconstructed FOAs, as well as the STFT and Mel distances using the default AuraLoss settings \cite{steinmetz2020auraloss}. For the SpatialVCTK dataset, we additionally report DistillMOS \cite{stahl2025distillation} and word error rate (WER). Spatial reconstruction metrics include average errors in azimuth, elevation, and angular distance for single-source scenes in the evaluation datasets.

It can be seen that our codec at 0.9kbps outperforms multichannel Opus at 24kpbs across all the datasets in most of the metrics, and at 32kbps on acoustic reconstruction metrics. Specifically, we can see that in the case of SpatialVCTK, which does not contain background noise and reverberation, our codec reconstructs the FOA signal with a mean angular error of 3.96°. Further, it achieves a DistillMOS score of 3.07 compared to 3.91 of the input FOA and a WER of 0.67 even though it is trained with small-scale speech data.

\begin{figure}[!h]
    \centering
    \includegraphics[width=0.8\linewidth]{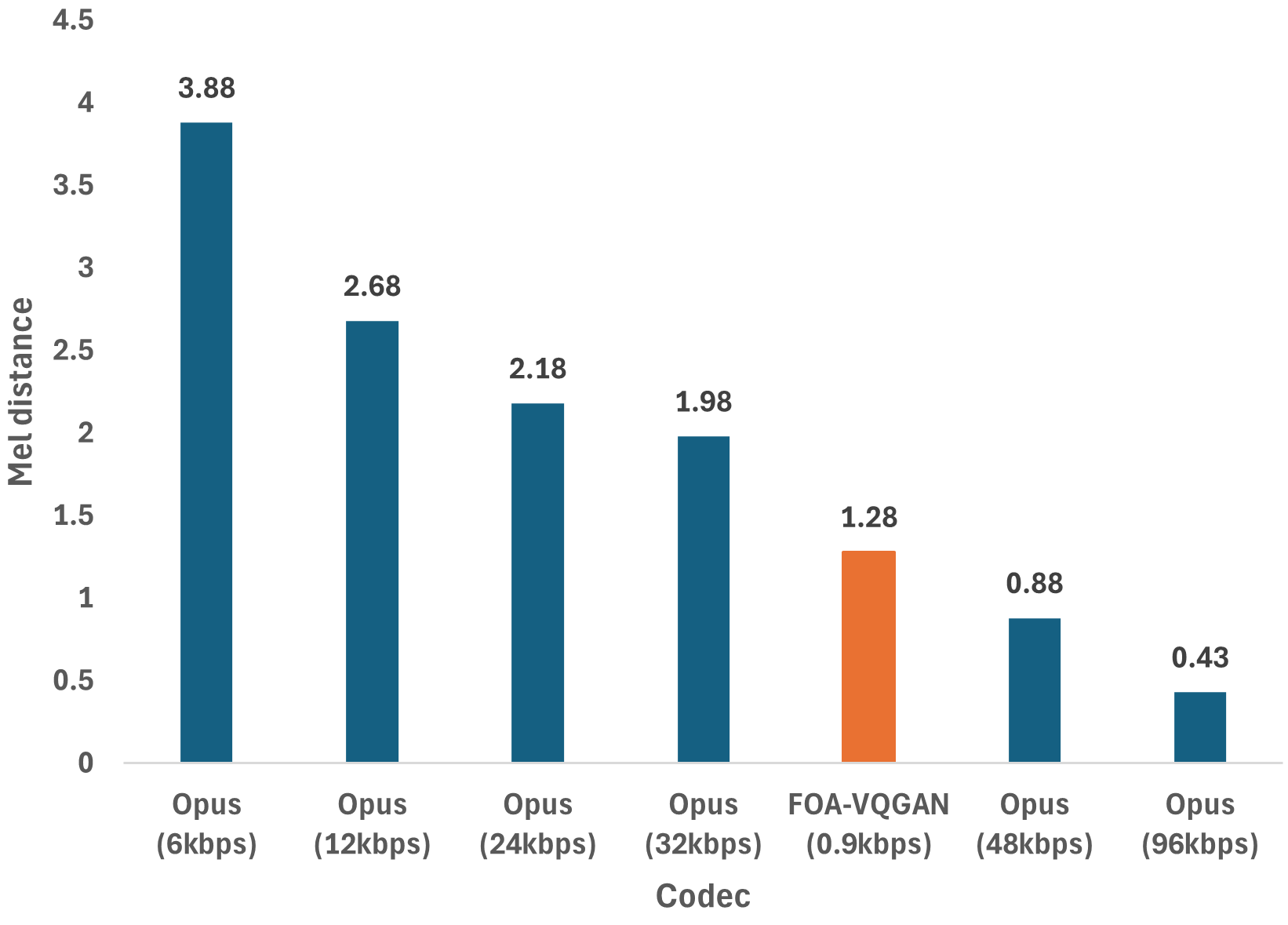}
    \caption{Comparison of acoustic reconstruction quality of our codec with multichannel Opus at various bitrates on the in-domain dataset.}
    \label{fig:mel_dist}
\end{figure}

\begin{figure}[!h]
    \centering
    \includegraphics[width=0.8\linewidth]{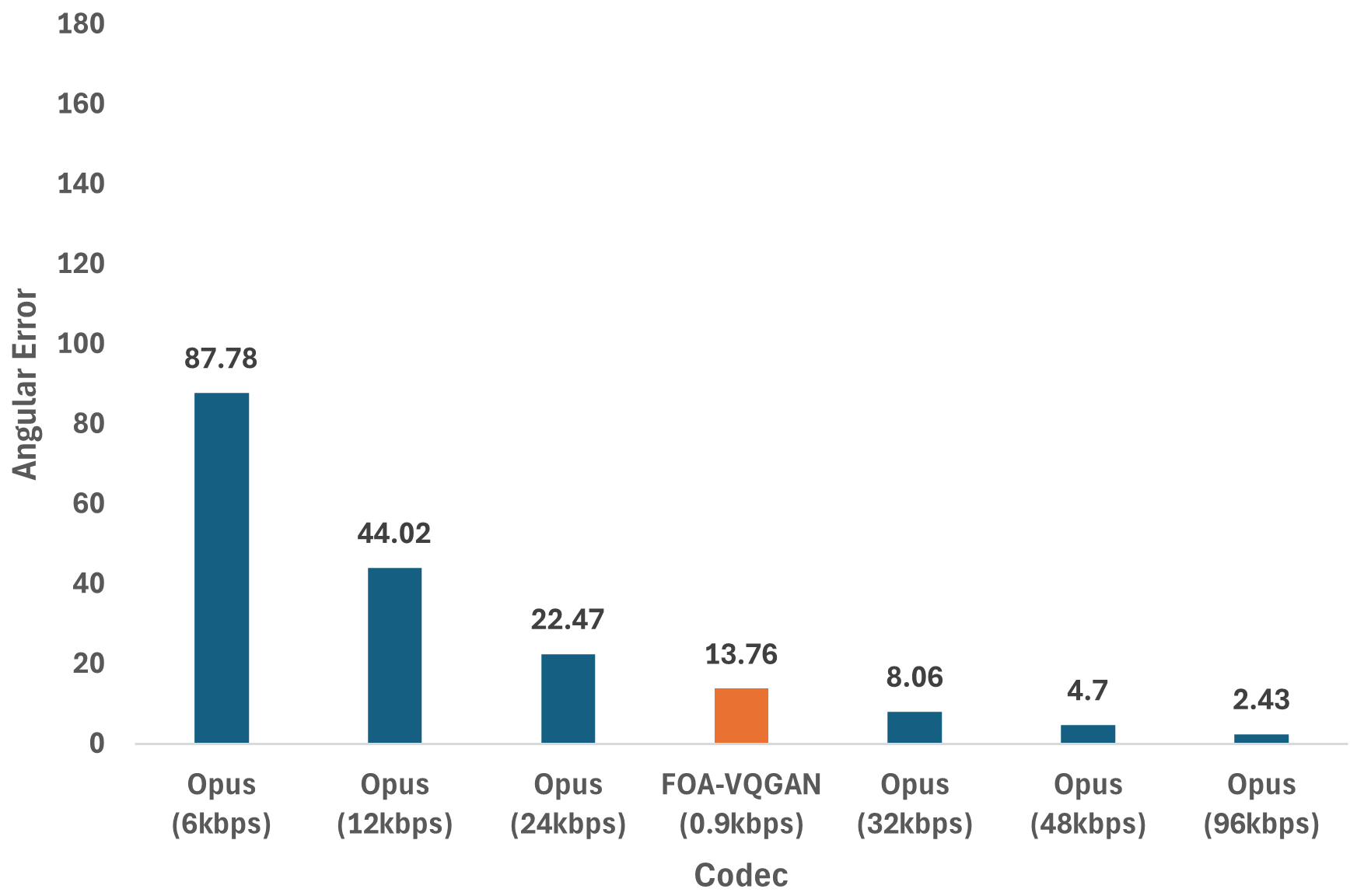}
    \caption{Comparison of spatial reconstruction quality of our codec with multichannel Opus at various bitrates on the in-domain dataset.}
    \label{fig:angular_err}
\end{figure}

In the in-domain data with unseen rooms and audio sources, our codec reconstructs with a mean angular error of 13.76 and STFT and mel distance of 1.60 and 1.28, respectively. Finally, in the MEIR dataset with real RIRs and sound sources, we achieve an angular error of 25.83°, showing the capability to transfer the learned knowledge into real recordings. FOA reconstruction examples from our codec are provided on the demo page.\footnote{https://partha2409.github.io/FOA-Tokenizer/}

Figures \ref{fig:mel_dist} and \ref{fig:angular_err} present a comparison of the acoustic and spatial reconstruction quality of our proposed FOA-VQGAN against the Multichannel Opus codec at different bitrates on the in-domain evaluation dataset. At 0.9 kbps, our codec achieves a mel distance comparable to that of Multichannel Opus operating between 32 kbps and 48 kbps. Likewise, in terms of spatial reconstruction measured by angular error, our codec performs on par with Multichannel Opus at bitrates between 24 kbps and 32 kbps. Similar trends were observed across the other reconstruction metrics and across all three datasets.

\begin{figure*}[!ht]
    \centering
    \includegraphics[width=1\linewidth]{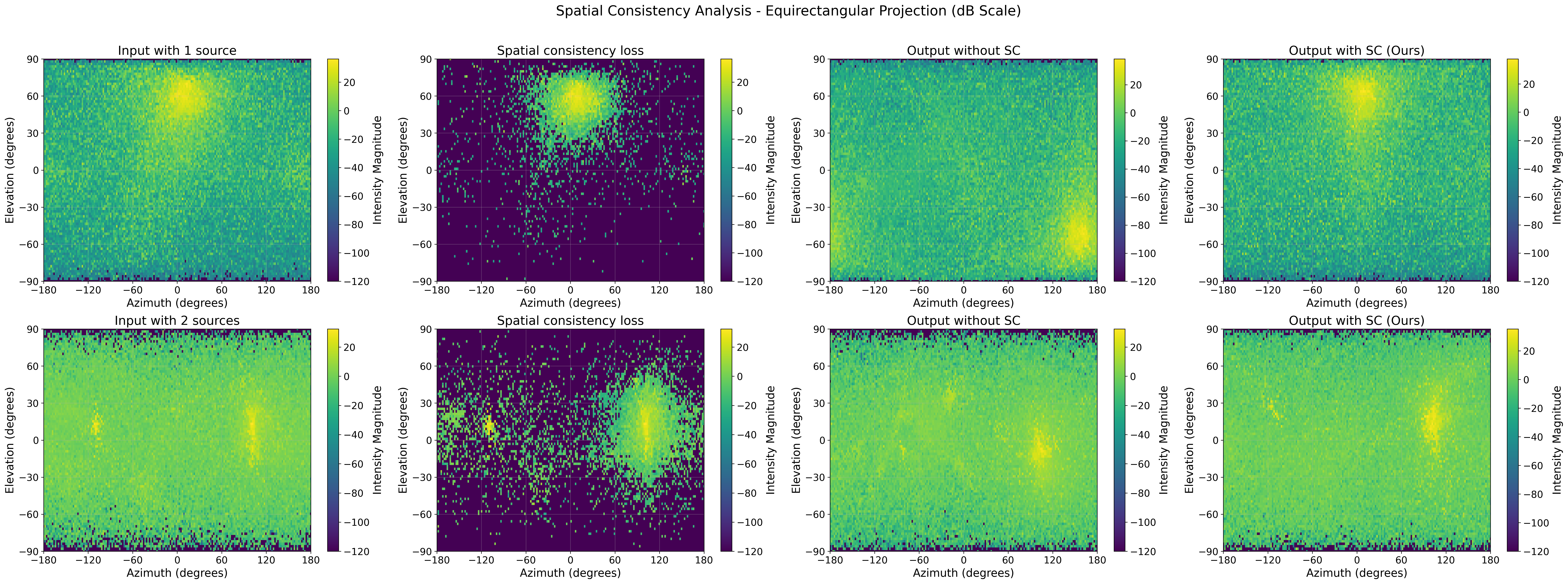}
    \caption{Spatial consistency visualization for FOA inputs with one and two sources, showing the intensity magnitudes of input, spatial consistency loss, and reconstructions without and with spatial consistency (SC) loss.}
    \label{fig:SC_loss_viz}
\end{figure*}

\vspace{-0.2cm}
\subsection{Impact of spatial consistency loss on spatial fidelity}

In Table \ref{tab:baselines}, we report the performance of the spatial audio codec trained without spatial consistency loss on the in-domain evaluation data. While its acoustic reconstruction metrics remain comparable to our proposed method, the mean angular error of 87.32° indicates a failure to preserve spatial properties at the compressed representation. We also evaluate a baseline that encodes each FOA channel independently using a pretrained WavTokenizer (4 x WavTokenizer), which likewise performs poorly in maintaining spatial information.

To illustrate the effect of the spatial consistency loss, Figure~\ref{fig:SC_loss_viz} shows time-frequency averaged intensity vector magnitudes for input FOAs and reconstructions from our spatial codec trained with and without the proposed SC loss. The SC loss intensity plot highlights how low-energy, highly diffuse regions are suppressed, guiding the model to preserve the spatial properties of the input in strongly directional time-frequency bins. It can be seen that for the model trained without SC loss, the sources are randomly spatialized in the reconstructions, whereas the proposed loss results in spatially faithful reconstructions.

\begin{table}[h]
\centering
\vspace{-6pt}
\caption{Performance of baseline models on the in-domain evaluation dataset.}
\vspace{0.2cm}
\resizebox{\columnwidth}{!}{%
\begin{tabular}{c|cccc}
\hline
\textbf{Model} & CLAP \(\uparrow\) & STFT Dist. \(\downarrow\) & Mel Dist. \(\downarrow\) & Angular Err. \(\downarrow\) \\
\hline
4 x WavTokenizer & 0.88 & 1.74 & 1.35 & 58.87° \\
FOA-VQGAN w/o SC & 0.92 & 1.61 & 1.30 & 87.32° \\
FOA-VQGAN (ours) & 0.92 & 1.60 & 1.28 & 13.76° \\
\hline
\end{tabular}%
}
\label{tab:baselines}
\end{table}
\vspace{-0.4cm}
\subsection{Evaluating quantized latents via SELD}

To evaluate the quantized representations, we performed sound event localization and detection (SELD). This task jointly measures the ability of the quantized latents to capture both sound events and spatial information, making it a strong probe of the acoustic and spatial content encoded by the codec. We conduct our experiments on the STARSS23 dataset \cite{NEURIPS2023_e6c9671e}, which contains real recordings of spatial audio scenes. To this end, we train a small SELD network on top of the quantized latents. Our network consists of 3 convolutional layers that downsample the time resolution of our quantized representations to match the label resolution of 100ms in the STARSS23 dataset and two fully connected layers to produce the SELD predictions in MultiACCDOA representation \cite{9746384}.  We compare our performance with the 
DCASE2023 SELD baseline\footnote{https://github.com/sharathadavanne/seld-dcase2023} model trained on per-channel mel spectrogram features and intensity vectors of the FOA signals. 

In Table \ref{tab:SELD}, we report the DOA-dependent F-score and the class-dependent localization error (LE) \cite{8937220} for both our model and the DCASE baseline. This F-score extends the standard F-score by requiring the estimated DOA to be within the threshold $\tau_{\text{DOA}}$ for a detection to count as a true positive. The LE is class-dependent, meaning the event class must be predicted correctly before the angular difference between the estimated and reference DOAs is measured. Our codec achieves comparable performance to the DCASE baseline at a $\tau_{\text{DOA}}$ of 45°, indicating that, while not sufficient for precise DOA estimations, the highly compressed representations still retain useful information for coarse spatial localization. It should be noted that our codec was trained only with stationary sources, while the STARSS23 dataset contains real recordings with moving sources that can further affect the performance. Hence, additional training of the codec on real spatial recordings (or real RIRs) could improve these results.

\begin{table}[h]
\centering

\caption{Performance of our model and the DCASE baseline on the SELD task on STARSS23 dev-test.}
\vspace{0.2cm}
\begin{tabular}{l|c|c}
\hline
\textbf{Model} & \textbf{F-score \(\uparrow\)} & \textbf{LE \(\downarrow\)} \\
\hline
DCASE Baseline ($\tau_{\text{DOA}}$ = 20°) & 29.9 & 22° \\
FOA-VQGAN ($\tau_{\text{DOA}}$ = 20°) & 11.1 & 37° \\
FOA-VQGAN ($\tau_{\text{DOA}}$ = 45°) & 25.3 & 37° \\
\hline
\end{tabular}
\label{tab:SELD}
\end{table}

\vspace{-0.5cm}
\section{Conclusion}
\vspace{-0.2cm}
In this work, we introduced FOA tokenizer, the first discrete neural spatial audio codec for first-order ambisonics. We extended the WavTokenizer to support FOA signals and proposed a novel spatial consistency loss to preserve directional characteristics in the reconstructions. Our spatial audio codec compresses FOA signals at 24 kHz into 75 tokens per second, corresponding to a bandwidth of 0.9 kbps. We evaluated the performance of our codec on simulated data with reverberant directional and diffuse sources, clean speech without reverberation, and FOA mixtures simulated with real RIRs. In all cases, our codec successfully reconstructed the audio with mean angular errors of 13.76°, 3.96° and 25.83°, respectively. Further, we presented preliminary experiments showcasing the use of the discrete representations for the sound event localization and detection task on the STARSS23 dataset. In the future, we plan to improve acoustic and spatial reconstructions by building new architectures designed to exploit interchannel relationships in ambisonic representations and explore generative spatial audio applications enabled by the discrete latent representations.

\vfill\pagebreak




\small
\bibliographystyle{IEEEbib}
\bibliography{strings,refs}

\end{document}